# Generalised Label-free Artefact Cleaning for Real-time Medical Pulsatile Time Series

*Xuhang Chen[1], Ihsane Olakorede[1], Stefan Yu Bögli[1], Wenhao Xu[1], Erta Beqiri[1], Xuemeng Li[3], Chenyu Tang[4], Zeyu Gao[5,6], Shuo Gao[2,3], Ari Ercole[7], Peter Smielewski[1],*

[1]: Brain Physics Laboratory, Department of Clinical Neurosciences, University of Cambridge, Cambridge, UK.

[2]: Hangzhou International Innovation Institute, Beihang University, Hangzhou, China.

[3]: School of Instrumentation and Optoelectronic Engineering, Beihang University, Beijing, China.

[4]: Department of Engineering, University of Cambridge, Cambridge, UK

[5]: Department of Oncology, University of Cambridge, UK

[6]: CRUK Cambridge Centre, University of Cambridge, UK

[7]: Division of Anaesthesia, Department of Medicine, University of Cambridge, Cambridge, UK

E-mail: xc369@cam.ac.uk

## Abstract

Artefacts compromise clinical decision-making in the use of medical time series. Pulsatile waveforms offer probabilities for accurate artefact detection, yet most approaches rely on supervised manners and overlook patient-level distribution shifts. To address these issues, we introduce a generalised label-free framework, GenClean, for real-time artefact cleaning and leverage an in-house dataset of 180,000 ten-second arterial blood pressure (ABP) samples for training. We first investigate patient-level generalisation, demonstrating robust performances under both intra- and inter-patient distribution shifts. We further validate its effectiveness through challenging cross-disease cohort experiments on the MIMIC-III database. Additionally, we extend our method to photoplethysmography (PPG), highlighting its applicability to diverse medical pulsatile signals. Finally, its integration into ICM+, a clinical research monitoring software, confirms the real-time feasibility of our framework, emphasising its practical utility in continuous physiological monitoring. This work provides a foundational step toward precision medicine in improving the reliability of high-resolution medical time series analysis.

## Introduction

Medical pulsatile time series, such as arterial blood pressure (ABP) and photoplethysmography (PPG) signals, provide vital insights into systemic physiology and clinical decision-making. These time series are continuously measured at high resolution either in hospital settings or ambulatory wearable healthcare applications, enabling the detailed analysis of dynamic waveforms and temporal trends to support the detection of clinical-relevant events[1]. However, these signals are often contaminated by artefacts from clinical interventions (e.g., blood sampling, line flushing) or patients' motion[2]. Such artefacts can distort signal characteristics, leading to misinterpretation of physiological events[3] or false alarms[4]. This risks alarm fatigue[3] and potentially inappropriate medical interventions at worst.

Over the past decades, various methods have been proposed to mitigate artefacts in medical pulsatile signals. Recent advances in machine learning artefact cleaning methods have shown superior performances compared to the traditional statistical and signal processing methods [5-7], particularly in addressing the challenges posed by non-stationary signals.

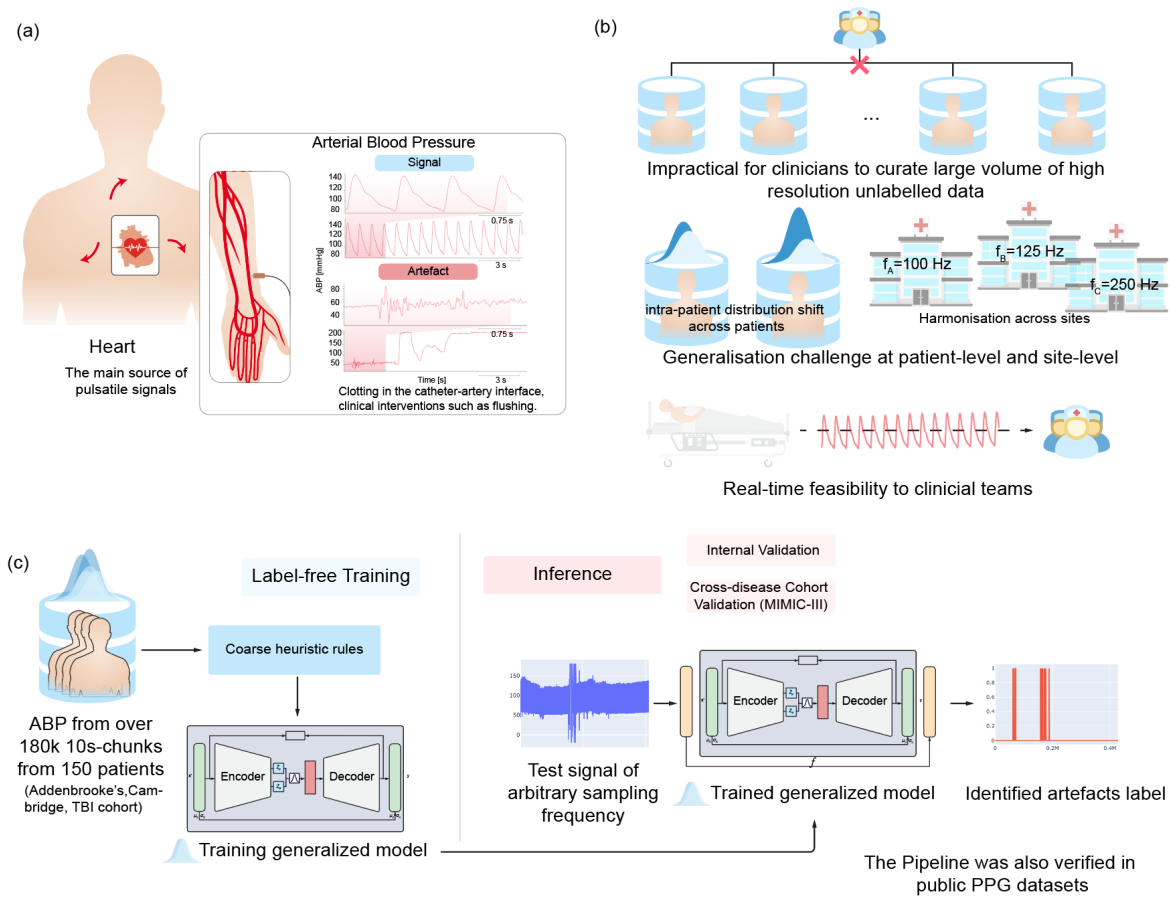

Figure 1 Overview of the challenges and framework for label-free medical pulsatile signal artefacts cleaning methods. a) Illustration of the origin of medical pulsatile signals and the arterial blood pressure (ABP) measurement. b) Key challenges in analyzing medical pulsatile signals: (i) the scarcity of artefact-annotated data (ii) generalization issue at the size- and patient-level (iii) lack of real-time feasibility as offline analysis is limited to retrospective analysis. c) The proposed framework for artefact detection. During training, data of 180,000 10-second samples from 150 patients, was used to train a generalized model. Coarse heuristic rules are injected to filter out the extreme values to improve the input data quality of our label-free method. During inference, the trained model can process signals that includes various generalization challenges, such as frequency inconsistency and patient-level distribution shifts. Additionally, we verified our effectiveness in both internal and cross-disease cohort MIMIC-III dataset. Also, this pipeline was verified in other medical pulsatile signals, such as PPG to demonstrate the generalizability.

Most contemporary methods are based on neural networks [3,8–13]. These approaches outperform traditional methods by leveraging their large number of parameters to perform end-to-end artefact cleaning on time series data.

Despite these advances, three major challenges (Fig. 1b) that hinder the broader adoption of the current methods can be identified. First, generalisation issues at both the site and patient levels limit the usage and performance of current methods. Site-level discrepancies (i.e., inconsistent harmonisation) are well-recognised[14], including device variations, protocols, and sampling frequencies across collection sites, which restrict the model application to specific sites where the data is collected. However, patient-level distribution shifts remain underexplored, yet degrade the model performance[15]. Distribution shift refers to the differences in statistical properties (e.g., mean, variance, non-linear and high order statistics) of variables of interest. It is particularly pronounced in medical time series signals and manifests itself at the intra-patient (e.g., ABP amplitude variations in physiological events) or inter-patient (e.g., individualised baseline ABP waveform) level [16]. Second, the scarcity of artefact-annotated data significantly limits the applicability of supervised learning methods. The absence of artefact-annotated data also highlights the urgent need for robust label-free methods in this domain. Third, the lack of real-time artefact cleaning power is a critical limitation of the current approaches. While retrospective analyses of medical time series aid research, real-time processing is essential for timely clinical decision support and false alarm reduction. Without real-time methods, the utility of artefact cleaning in dynamic, high-stakes environments remains restricted.

In this work, we address these gaps by developing a generalised label-free framework, GenClean, with the inclusion of feasible real-time implementation (Fig. 1). To this end, we first conducted quantitative experiments to examine patient-level generalisation. Subsequently, we validated our method through cross-cohort experiments (i.e., train on one cohort and test on another) on medical pulsatile time series, ABP and PPG signals, showcasing its effectiveness for artefact cleaning. Leveraging our label-free training method, we utilise 180,000 10-second (500 hours) ABP samples to train our model, extracted from traumatic brain-injured patients treated in Addenbrooke's Hospital, Cambridge. We further validated our method in more challenging generalised scenarios, such as cross-disease cohort settings, validated on MIMIC-III patients. Additionally, we integrate our method into ICM+, a clinical research monitoring software[17] (Cambridge Enterprise Ltd, Cambridge, UK), to verify its real-time feasibility in clinical workflows. To the best of our knowledge, this is the first artefact-cleaning method that demonstrates the generalisation ability across different clinical sites and varying harmonisation levels of medical signals. We aim to provide a practical generalised label-free method that can assist precise medical signals analysis in the wide medical environment.

## Results

### Clinical Datasets

We used both ABP and PPG datasets to evaluate our model. The primary ABP dataset (120 Hz) comprised data from patients (n=160; ≥18 years) with trauma brain injury (TBI), sourced from Brain Physics Database, Addenbrooke's Hospital, Cambridge, under ethical approval REC 23/YH/0085. Of these, data from 150 patients were allocated to the training set and validation set, while data from the remaining 10 patients were reserved as an internal held-out set, which was balance labelled (non-artefactual: artefactual = 1:1). We segmented both datasets into 10-second samples to manage the data efficiently, a duration that minimises computational and data-loading burdens while maintaining alignment with clinically relevant metrics, such as the pressure reactivity index (PRx)[18], a key metric used for continuous cerebral autoregulation monitoring in TBI. Overall, this process yielded 180,000 10-second (500 hours) samples from the training and validation cohort and 1,000 balance-labelled 10-second samples from the internal test cohort.

To evaluate the generalisability of our model to a cross-disease cohort, we incorporated ABP signals from five patients retrieved from the MIMIC-III Waveform Database[19] as a cross-disease cohort validation set. This cohort encompassed a broader range of patients from intensive care units (ICUs). Moreover, photoplethysmography (PPG) datasets, including DaLiA (n=15; 1,837 10-second samples) and WESAD (n=15; 8,664 10-second samples)[20,21], were used to demonstrate further the validity of our model on other types of medical pulsatile signals. A full description of our methods and the training details is provided in the Methods.

### Distribution Shift Exploration

We investigated inter and intra patient distribution shift issues in the non-artefactual part of the internal held-out set. Inter patient distribution shifts can be observed via differences in statistical summaries (e.g., median, 1 and 1.5 interquartile range, Fig. 2a). All ten patients showed different mean, median and quartile values. For example, the median of PT 6 was close to 90 mmHg and that of PT 3 was nearly 75 mmHg. This was confirmed using a two-sample Kolmogorov-Smirnov test with the Bonferroni correction for each pair of patients, which showed a significant difference between each pair of patients (p<0.05). Intra patient distribution shifts can be characterised by peaks and tails in the histogram (Fig. 2b), particularly the multi-peak distribution patterns. Each peak represents a distinct range of frequently occurring ABP values. These peaks likely correspond to different physiological states experienced by the patient over time. For instance, the four-peak pattern in PT 10 suggests that the patient transitioned through multiple physiological states or events during monitoring, showing the intra patient distribution shifts. Fig. 2c employs t-distributed stochastic neighbour embedding (t-SNE) to project 10-second signal samples into two dimensions, visualising qualitative distribution shifts at the sample level for selected patients. PT 1, 2 and 3 formed distinct groupings, suggesting distributional differences among these patients, while data from PT 4 and PT 5 were scattered across multiple clusters, suggesting overlapping distribution characteristics. These clustering results reveal the intricate distribution patterns within the ABP signal. Overall, these observations demonstrate distribution deviations in patient ABP data.

Building on these observations, we evaluated the performance of our previously proposed label-free method[8] in the face of the distribution shift challenge on the test set. In Fig.2d, the $10 \times 11$ generalisation matrix visualises the metrics (accuracy and F1 score) of training on patient $i$ and testing on patient $j$ where each row indicates training on the same patient, while the last column shows the results from our generalised label-free method trained on our training set. The diagonal line of both matrices manifests training and testing on the same patient, while the other area indicates cross-patient validation. The F1 score matrix was similar to the accuracy matrix, with slightly lower F1-score values (deviation of up to 0.3 - Train 1 Test 3). Our generalised model exhibited excellent performance with at least 90% accuracy and 0.89 F1-score even in the distribution shift settings, outperforming the previous method across all the patients. In the previous method, seven out of ten patients achieved satisfactory performance (>70%) in the generalisation test. Additionally, the diagonal line (self-test) tends to exhibit higher accuracy, however, this trend was not consistently observed in our data on the accuracy matrix (accuracy of 0.80 for Train 1 Test 10 vs. 0.60 for Train 1 Test 1). One of the patients (Test 5) consistently exhibited lower performance in both metrics, with no value above 0.6. This distribution shift observation highlighted our generalised model's superior performance (Fig. 2d) in dealing with this complexity where previous methods often struggled[8].

**Generalisation on Cross-cohort**

We further evaluated the generalisation performance of our models in data sets with different sampling frequencies of signals and another cohort of patients. Fig. 3b displays our model performance on accuracy and processing time in a range of common sampling frequencies (50, 75, 100, 120, 125, 150, 175, 200 and 240 Hz). The result indicates no notable performance degradation across varying sampling rates. Regarding the time for artefact cleaning, it takes ~19 ms for 120 Hz, which is the frequency that our model trained on, with the longest processing time reaching ~26 ms (150 Hz). This result confirms the robustness of our method in handling diverse frequencies.

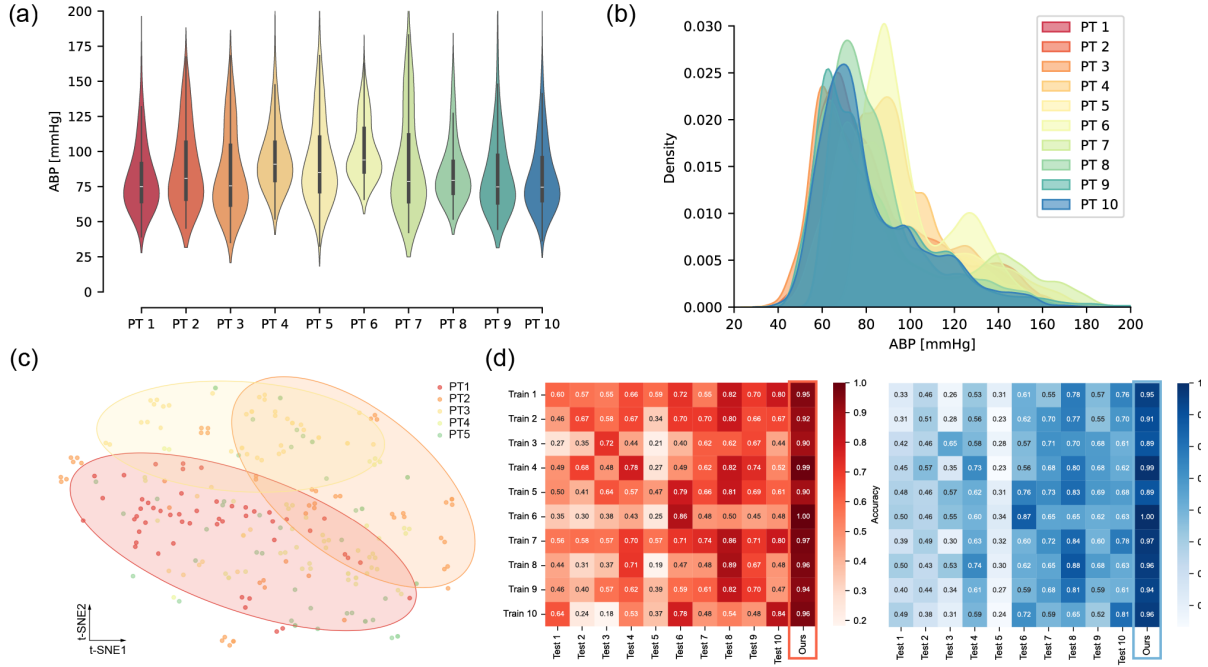

Figure 2. Visualisations of data distribution, generalisation, and model performance. (a) Violin plots of non-artefactual ABP signal values for 10 patients, illustrating the distribution of high resolution arterial blood pressures. Clear inter-patient variability is observed, reflecting physiological and pathological differences among patients. (b) Density plots of ABP distributions for the same 10 patients. The overlapping regions highlight intra-patient similarities, while distinct peaks in certain patients suggest inter-patient differences (unique physiological characteristics in each patient). (c) t-SNE visualization of the first five patients, selected to enhance clarity and reduce visual clutter. The distinct clusters show three patients with clear distribution shifts, while the other two exhibit overlap with these clusters, indicating potential patient-specific morphological variability. (d) Generalisation matrices comparing accuracy (left) and F1-score (right) for cross-patient training and testing (train on patient i, test on patient j) in the previous label-free method. The last column is the performance of our generalized label-free method. The result highlights the model's generalisation capability, with better performance seen in patients with consistent data distributions.

To test the model's generalisation further, we conducted a cross-disease cohort validation experiment on randomly selected ABP signals from the MIMIC-III waveform database. Here, the model trained on our internal set (TBI cohort, 120 Hz) was directly applied to MIMIC-III data (general ICU cohort, 125 Hz) to verify the artefact cleaning effects on out-of-distribution patients and inconsistent data collection standards. An accuracy of 95.6% was obtained, with further quantitative results available in Table 1. Fig.3a displays a data overview of one patient and the detected label from our model, with three expanded illustrative pulse waveform sections, called cases. Cases 1 and 3 show very reasonable model performance on pulses with different morphology and statistical properties (e.g., mean and standard deviation), correctly identified as artefact-free. Case 2 demonstrates a distorted trace, which our model identifies correctly as an artefact, with the decoded waveform being very different from the original. These three cases also illustrate the intra-patient level distribution shift.

Table 1 Specific performance of the proposed artefacts cleaning method in continuous monitoring ABP and PPG dataset.

| Metrics | MIMIC-III (ABP) | WESAD (PPG) |
| --- | --- | --- |
| Accuracy | 95.6% | 85.8% |
| Specificity | 97.2% | 86.7% |
| Sensitivity | 94.3% | 85.3% |
| Hypertension pulses before cleaning | 2134 | N/A |
| Hypertension pulses after cleaning | 1661 | N/A |
| Reduced Events proportion | 22% | N/A |

To shed light on the potential clinical relevance, we analysed the impact of artefact cleaning on the detection of hypertension events. Hypertension, commonly defined as systolic blood pressure exceeding 140 mmHg or diastolic blood pressure exceeding 90 mmHg[22], can be directly determined from ABP signals and can provide critical insights for the metabolism control of patients. We applied our model to five patients of the MIMIC-III dataset and counted the reduction in the detected number of hypertensive events after cleaning. The results (Table. 1) showed a 22% reduction in hypertensive events count, highlighting the model's practical value.

Furthermore, to explore our model's generalisation in other medical pulsatile time series, we used PPG signal as a candidate for testing[12], as artefacts cleaning on PPG signal is a challenging task, particularly when the data is collected on wearables and is more sensitive to motion artefacts. We trained our model on the DaLiA dataset and tested it on the WESAD dataset to verify the model performance on a completely different signal from the one it was originally developed for. Our results (Table. 1) revealed an 85.8% accuracy in the PPG dataset, demonstrating the feasibility of transferring the approach to different medical signal modalities.

**Model Architecture Analysis**

We conducted ablation studies (i.e., modifying the components of our method) and compared the results to baseline methods (Fig. 3d). The performance of five evaluation metrics was exhibited throughout our ablated models and baseline methods. These include one-class support vector machine (OCSVM)[23], a classic label-free method, ResNet1D[24] and XGBoost[25], two famous supervised learning methods proven to achieve decent outcomes for artefacts detection, and our three ablated models (named w/o) without specialised design for generalisation (i.e., distribution shifts). Our model (labelled 'Ours' in the figure) outperformed all the models, even including the supervised learning methods in all five metrics. Training over a large patient dataset (150 patients) without addressing generalisation issues significantly degrades performance. This result demonstrates the validity of our model design and the superior performance as a label-free method.

**Latent space representation**

We analysed the impact of our latent dimension from the perspective of model design. The latent dimension in Fig. 4 refers to the size of the compressed representation (i.e., the latent vector) of waveforms, which represents the space in which the input signal is encoded and later reconstructed by the decoder. This latent vector serves as an embedding of the compressed physiological information of the medical pulsatile time series included in our model. Fig. 3c illustrates the impact of varying the latent space dimension in the variational autoencoder on the F1 score, as this metric is stricter than Accuracy as shown in Fig. 2d. Our results indicated that a dimension size of 20 provided an optimal balance, with the highest score observed when mean squared error (MSE) was used as the reconstruction error evaluation metric. Deviations from this dimension size resulted in depleted performances.

To evaluate the effectiveness of the representations learned from our model, we conducted a detailed qualitative analysis of the latent vectors. Two widely used dimension reduction methods, t-SNE and Uniform Manifold Approximation and Projection (UMAP), were applied to visualise and disentangle the learned latent representations. Before applying our model, Fig. 3e shows the results from our internal held-out set, where points of both categories are intermixed within the labelled blue region. After applying our model, Fig. 3f displays the learned latent vector maps. The latent space demonstrated improved separation between the two categories. This suggests that the latent space learned by our model effectively captures the underlying data structure, enhancing its ability to distinguish between artefactual and clean signals.

| Table 2 Real-time feasibility metrics of our artefacts cleaning method measured in ICM+ | | |
|---|---|---|
| Metrics | Value | Description |
| Processing Time | 39 ms | Average time to process a 10-second segment, ensuring real-time feasibility. |
| Model Computation | 3.2 mFLOPS | Computational cost per segment |
| CPU Utilization Increment | +6% | Average CPU load during processing, compared to baseline system utilization. |
| Memory Usage Increment | +139 MB | Additional memory used for model inference during the experiment. |

**Real-time feasibility**

Most of the previous methods were demonstrated in offline analysis in retrospective datasets [8,26–29]. Real-time artefact cleaning capability is however essential for effectively suppressing false alarms and providing support for clinical decision-making. To evaluate the real-time feasibility of the method, we applied our GenClean method to a simulated normal monitoring data stream using ICM+, a clinical research monitoring software[17].

As shown in Fig. 3g, the left panel displays a long-period monitoring graph, while the right panel provides fine-grained patient data and the decoded waveforms produced by our model. Table 2 summarises the quantitative computation overheads during the running period of our method as a plugin. The increments of running our model on the central processing unit (CPU, +6%) and memory (+139 MB) were within a reasonable range and the processing time negligible, making this approach feasible even when using low specifications hardware for bedside data collection/processing. The computation cost measured in floating-point operations (FLOPs), was significantly lower (3.2 mFLOPS) than deep neural networks like ResNet1D (3 GFLOPS). These findings confirm the practical applicability of our approach in real-time clinical environments.

## Discussion

Building upon insights from DeepClean[8], we developed GenClean, a state-of-the-art, generalised, label-free method for artefact cleaning and demonstrated its real-time feasibility on clinical monitoring software running on CPU. Our framework, trained on 180,000 10-second samples of data collected at Addenbrooke's Hospital, integrates a generalisation design and a label-free training strategy. Despite challenges such as site-level harmonisation issues and cross-disease cohort variability, our model achieved over 90% artefact recognition accuracy on the MIMIC-III dataset. When applied to a PPG signal, a widely used non-invasive medical pulsatile signal, our method delivered robust results with an accuracy exceeding 85% on publicly available datasets.

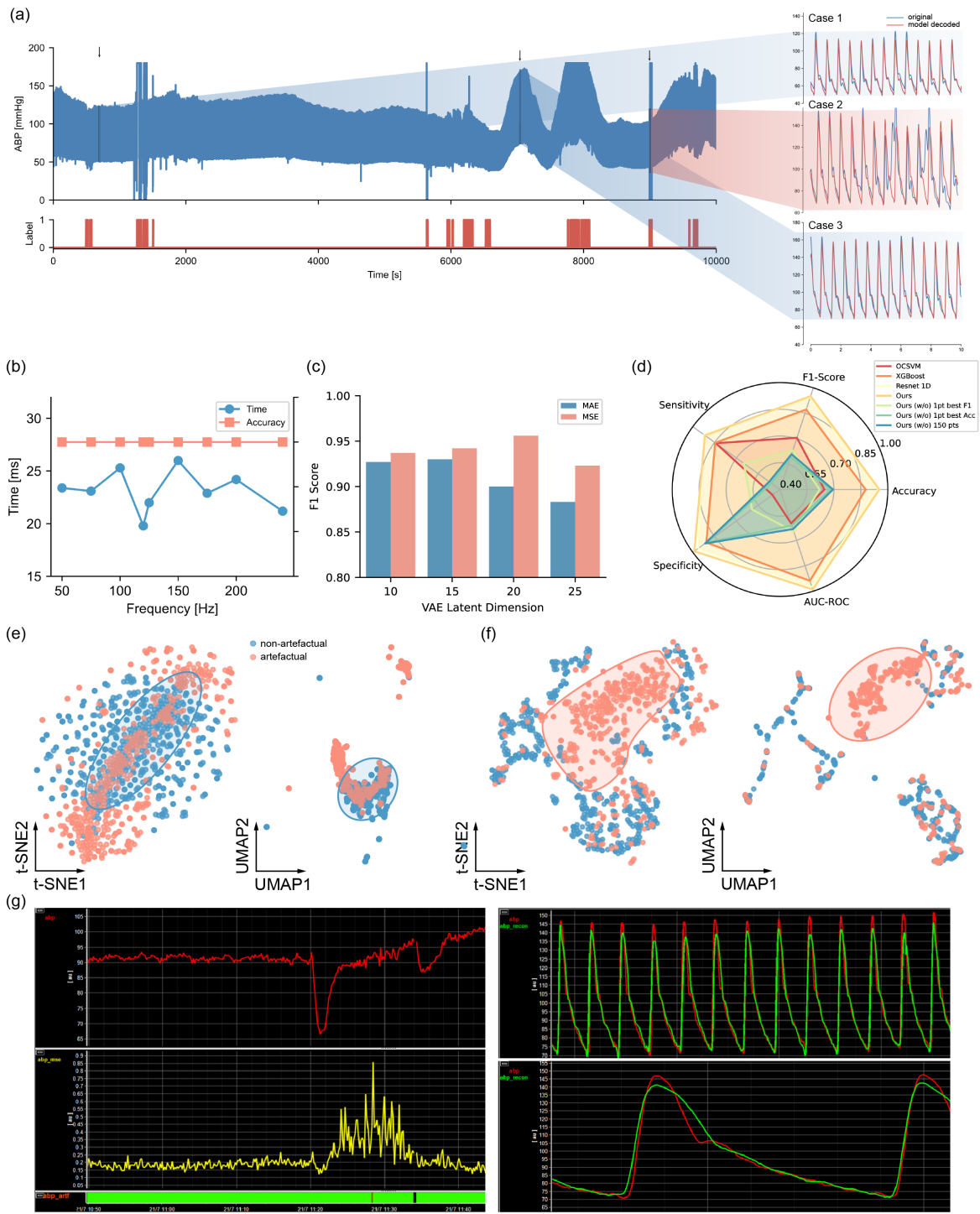

Figure 3. Model performance visualisations and real-time feasibility. (a) Example of processing of an ABP recording from the MIMIC-III database. The left panel displays a section of the ABP waveform including dynamic changes and artefacts, with the model generated artefacts markup (below). On the right, Case 1 and Case 3 showcase different physiological stages with correctly reconstructed ABP waveforms. Case 2 represents an artefact instance, where the model successfully identifies the section as anomalous, with a clear discrepancy between the original and reconstructed signals. (b) the accuracy and processing time of our method at various input sampling rates (120 Hz is the frequency our model trained on). (c) Impact of the latent space dimensionality and error evaluation metrics (MSE and MAE) on model performance. (d) Comprehensive comparison of model performance using a radar plot, including baseline models, alongside our model variants with ablation settings, highlights our model's superior generalization and accuracy across metrics. (e) and (f) Visualisation of data and latent space distributions using t-SNE and UMAP. The left panel illustrates the original data with significant overlap in the blue-highlighted region, indicating limited separability. The right panel displays the transformed latent space, where clear clustering emerges, demonstrating the effectiveness of the model in separating physiological categories post-processing. (g) Left: ICM+ screenshot showing example of monitored signals, with the error (abp_mse) between reconstruction and the origin in the middle, artefact annotation panel (abp_artf) displayed at the bottom. Right: the real-time reconstructed pulse wave, reflecting our explanatory panel.

The impact of artefacts in medicine was historically underestimated due to the predominance of low-resolution datasets [26,30–32], and sustained poor support of the medical device industry for high-resolution data integration and extraction of waveform features[33]. Artefacts in current high-resolution data significantly distort waveform and trend analyses, impacting subsequent clinical assessments. Previous artefact cleaning studies in this field often focused on small-scale, private datasets [28,34,35], or one single patient (DeepClean)[8], and mostly relied on supervised methods [28,34,36]. A major challenge in earlier studies was the lack of attention to generalisation including at the patient-level (the distribution shift), which degraded the models' performances and limited the usage on large populations. Our work first investigated this issue in ABP signals and demonstrated the improvements achieved through our proposed methods. Notably, the original version of this method[8] still performed relatively well in 4 out of the 10 tested patients, potentially explaining why this issue has been overlooked in the previous study. We validated our trained model in the MIMIC-III using different data sampling frequencies as a feasibility exploratory to handle generalisation issues across sites. By resolving both patient- and site-level generalisation challenges, our approach effectively utilises our 180,000 samples and achieves competitive performance against supervised methods. Our method can mitigate the risks associated with distribution shifts, improve the data quality for medical models and reduce the likelihood of the "garbage in, garbage out" problem.

A key innovation of our work lies in its real-time feasibility, a critical requirement for bedside monitoring. While many existing methods have been developed and validated using retrospective datasets[8], real-time detection is essential for continuously tracking patient status and enabling timely intervention. Our framework addresses this issue by integrating our model as a plugin into a clinical research monitoring software to validate the feasibility of this approach. Due to stringent privacy restrictions in clinical settings and limiting possibilities of streaming data in real-time to high-performance computing infrastructures, real-time feasibility must be achieved on CPU-based bedside hardware, which constrains many existing methods[9,37]. The reasonable computation overheads of our approach demonstrate its feasibility for application in a modest hardware requirement setting. Furthermore, we also demonstrated that artefact cleaning significantly reduced false hypertension events, which might otherwise trigger frequent monitoring device alarms and even lead to alarm fatigue among clinical staff [38]. Our framework maximises the value of artefact cleaning, potentially improving the accuracy of real-time clinical metrics such as the pressure reactivity index[18]. These improvements are critical for advancing precision medicine by providing more accurate reflections of patient physiology and supporting individualised care.

In addition, our framework benefits significantly from the variational autoencoder (VAE) backbone. Previously widely used in image generation[39], the VAE's effectiveness in this domain has been validated in DeepClean[8]. Within our framework, the VAE model provides three distinct advantages. First, our design allows artefact detection to be more effectively achieved through the decoder's outputs. By generating non-artefactual segments based on input signal features, deviations (e.g., Fig. 3a, cases 1–3) can be directly observed to assess signal integrity and potentially recover contaminated sections[8]. This approach is more intuitive and compelling than post-hoc methods, such as Grad-CAM[40] or SHAP[41], and more interpretable than black-box feature engineering algorithms. Second, compared to similar methods[26], our robust VAE backbone can obtain more standardised and transparent metrics for artefact identification among patients, which avoids the use of subsequent complex machine learning methods on latent vectors to identify artefacts. Also, unlike autoencoder methods, the probabilistic modelling of VAEs provides continuous probability modelling in the latent space, ensuring a smooth and continuous representation of the data. This continuity allows for a more structured latent space, enabling meaningful dimensionality reduction, as demonstrated by t-SNE or UMAP latent space visualisations. Finally, the latent vector may add value to the understanding of patients' states, as the essence of understanding artefacts lies in identifying and distinguishing features of normal signals. Thus, continuous monitoring facilitates the analysis of temporal trajectories of patient status which could potentially highlight patient-specific conditions and distributions. To achieve this, potential spatial transitions corresponding to different clinical states or labels may need to be identified.

In selecting medical pulsatile signals, we focused on two representative cases, invasive ABP, the clinical gold standard for continuous blood pressure monitoring, and PPG, one of the most used non-invasive pulsatile signals in consumer healthcare electronics. Testing our model on these two signals demonstrated strong performance, indicating its potential applicability to a broader spectrum of medical pulsatile signals. Medical pulsatile time series are generally driven by cardiac activity, characterised by consistent waveforms, and governed by well-defined physiological principles. This inherent stability ensures the presence of reliable non-artefactual sections, which our method leverages to learn physiological patterns while addressing generalisation challenges across samples. By doing so, our framework enables large-scale label-free training and holds promise for application to other medical signals. Furthermore, our work does not diminish the value of alternative approaches, such as supervised or statistical methods. On the contrary, we advocate for a multi-stage methodology: our label-free framework can be employed for large-scale artefact cleaning during the initial stages of data processing, while identified labelled cases from this process can further inform supervised training. We believe that semi-supervised and supervised approaches remain highly complementary and encourage their continued exploration and development alongside label-free methods to advance this domain.

Despite its strengths, our model has certain limitations. First, the training dataset consisting of 150 adult TBI patients may not be diverse enough to generalise well to other populations, such as paediatrics. While our approach mitigates distribution shifts, the effectiveness still depends on dataset representativeness. Expanding the dataset to a broader population across ages can facilitate its applicability. We acknowledge that collecting these larger datasets and addressing the site-level generalisation problems within the data (e.g., differences in collection protocols and device configurations) still requires a concerted effort from the research community. Second, as a probabilistic model, the VAE may struggle with out-of-distribution cases not covered in the training set, such as cardiac arrhythmias, as the model learns to reconstruct common patterns from the training data. When faced with unseen physiological variations, the model may incorrectly assign them to the closest learned distribution in common cases. Addressing those special cases may require targeted solutions. Third, implementation for real-time processing of signals from patient monitors requires more attention. Future efforts will focus on deploying and testing the framework in bedside environments to assess its practical impact on patient care.

Our study highlights the transformative potential of label-free methods in artefact detection and cleaning for medical pulsatile signals. By tackling challenges like artefacts, distribution shifts, and frequency variability, we provide a robust and generalisable framework for real-world digital medicine applications. The open-source nature of our framework promotes accessibility and collaboration, paving the way for advancements in medical time series analysis. These contributions mark significant progress towards ensuring high-quality data, an essential step for further robust data-driven precision medicine and individualised patient care.

# Methods

## Data Preparation

We used ABP data sourced from the Brain Physics Research Database as an internal training and test set and conducted a cross-disease cohort validation on the MIMIC-III datasets. The Brain Physics database ABP data (n=160) was measured by arterial line (Baxter Healthcare, Deerfield, Illinois) inserted into the radial or femoral artery and recorded at a frequency of 120 Hz using ICM+ software in patients with traumatic brain injury admitted into the Neurocritical Care Unit, Addenbrooke's Hospital, Cambridge, UK (2011-2019), requiring intracranial pressure monitoring. ABP was monitored through radial or femoral arterial lines connected to pressure transducers (Edward Lifesciences, Irvine, CA). For the training set, each patient contributed 1,000 10-second samples, randomly sampled from their monitoring period, while 200 samples per patient were allocated to the validation set. This resulted in a total of 180,000 10-second samples. Each patient provided 100 balanced 10-second samples, amounting to a total of 1,000 samples (~166 minutes), ensuring sufficient representation of individual physiological variability in the dataset. The cross-disease cohort, MIMIC-III, was collected from critical care units of the Beth Israel Deaconess Medical Center in Boston, USA (2001-2012). ABP was collected by an HP CMS (Merlin) monitoring device at the sampling frequency of 125 Hz [19]. We randomly selected five patients with three-hour sections as a validation experiment due to the scarcity of patients with ABP data. However, it is worth noting that for the inference, our framework can process them without segmentation, as illustrated in Fig. 3a. For additional medical pulsatile time series validation, we utilised a publicly available PPG dataset[12], commonly utilised in healthcare applications. The DaLiA dataset (n=15; 1,837 10-second samples) included recordings from 15 participants performing daily tasks such as walking, cycling and driving. The WESAD dataset (n=15; 8,664 10-second samples) included recordings from 15 participants in different emotional states such as neural, stressed, and amused. The PPG signals were segmented into 10-second samples to align with the input shape of our model. Following the previous work from Chen et. al[12], we employed the DaLiA dataset for label-free training and the WESAD dataset for testing.

## Heuristic rules

Medical pulsatile time-series signals, such as ABP and PPG, provide rich physiological information and exhibit high interpretability compared to other biomedical signals (e.g., EEG). We applied heuristic rules based on domain knowledge to filter out extreme artefactual or non-physiological signals (e.g., negative values, flat lines) that are unsuitable for clinical decision-making. For arterial blood pressure, we considered a normal range to be from 0 mmHg to 300 mmHg, and the peak-to-peak value to be above 15 mmHg. For the PPG signal, we applied a frequency filter to suppress the frequencies outside of the 0.5 Hz – 3 Hz range. Segments where more than 30% of the total signal power falls outside this range were classified as artefacts[42]. These criteria can enhance the quality of data for training label-free variational autoencoders to learn physiology while minimising the impact of extreme artefacts.

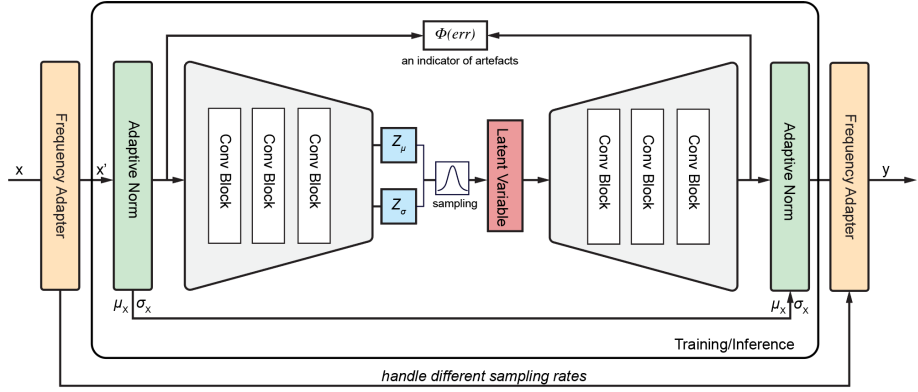

Figure *4 Our artefact cleaning framework with two designed modules, frequency adapter and adaptive norm to handle the generalization challenges at site-level and patient-level, and the core variational autoencoder backbone to provide probabilistic modelling of non-artefactual medical pulsatile time series.*

## Artefact Cleaning Framework

To address the challenges posed by diverse datasets in artefact cleaning, we propose a unified framework (Fig. 4) combining the frequency adaptation, variational autoencoder (VAE)-based backbone, and adaptive norm module. This design ensures compatibility across varying sampling rates for site-level generalisation, robust representation of clean waveforms, and effective handling of patient-level distribution shifts.

We apply the variational autoencoder to learn the probabilistic representation of clean waveform which constrains the learned normal waveform, leveraging a one-dimensional convolutional structure in the encoder and decoder. This architecture focuses on local signal features, aligning with clinical insights into pulsatile signal characteristics. This design considers the trade-off between the feasibility of real-time implementation, physiological domain knowledge, and algorithm robustness. The loss for training is selected as the standard evidence lower bound objective loss[39]:

$$L = E_{q_{\varphi(z|x)}}[\log p_{\theta(x|Z)}] - D_{KL}(q_{\varphi(z|x)}||p_{\theta}(z))$$

where $p_{\theta(x|z)}$ is the encoder, $q_{\varphi(z|x)}$ is the decoder. The size of the input $x$ is 1200 (120 Hz by 10 seconds), and the size of the latent variable $z$ is 20.

The Frequency Adapter is designed to manage high-resolution data from multiple sites (e.g. 125 Hz for the MIMIC-III vs. 120 Hz for the training set). Input signals are resampled to the target frequency the model requires, and outputs are resampled back to the original frequency for downstream analysis. This reversible process preserves the original temporal structure while ensuring compatibility with the model's training pipeline.

In our study, the sampling frequency is provided in the HDF5 data stream from ICM+ software, and a tool for extracting it is available in our code. Building upon this adapter structure and estimated frequency design, our module can handle a direct one-dimensional signal input and return the exact same length decoded results and corresponding label (Fig. 3a), enabling convenient usage.

Furthermore, to address the patient-level distribution shift issue, we applied reversible instance normalisation[43] within a large group of patients. Considering a sample $x_k \in R^C$, the mean and variance at the sample-level are,

$$E[x_k] = \frac{1}{C}\sum_{j=1}^{C} x_{k,j}, \qquad Var[x_k] = \frac{1}{C}\sum_{j=1}^{C}(x_{k,j} - E[x_k])^2$$

Let denote the original self-supervised network as $M(\,\cdot\,;\omega): R^C \to R^C$, which takes a $R^C$ dimension vector $x_k$ and return the same dimension output $y_k$ with $y_k = M(x_k;\omega)$. Currently, with the reversible instance normalisation, the input of $M(\,\cdot\,;\omega)$ becomes,

$$\hat{x}_k = \frac{x_k - E[x_k]}{\sqrt{Var[x_k] + \epsilon}}$$

Where $\epsilon$ is a small value to avoid division by zero. To restore the original scale after processing, the output of the model is transformed back as,

$$\hat{y}_k = \sqrt{Var[x_k] + \epsilon}\, M(\hat{x}_k;\omega) + E[x_k]$$

The symmetric structure of this adaptive normalisation layer within the network is performed at each iteration during training. This incorporated method preserves this relevance and ensures our network learns the meaningful part from signals.

Finally, artefacts are identified based on reconstruction error between the input signal and the decoder output. We empirically used the mean squared error (MSE) as a metric (Fig. 3c), and the artefact threshold was empirically set as the 90th percentile of our validation set. Segments exceeding this threshold are classified by the algorithm as artefact contaminated. This approach ensures robust artefact detection. To evaluate the performance of artefact identification, based on the artefactual sample and our experts-labelled data, we employed a comprehensive set of metrics, including accuracy, sensitivity, specificity, F1-score, and the area under the receiver operating characteristic curve (AUC-ROC) (Fig. 3d). We consider our metrics to be representative enough due to our balanced test set setting.

**Training methods**

The training method is a critical aspect of ensuring the convergence and correctness of the generative model. First, the vector used for calculating the loss function was carefully selected. The reconstruction loss (the first part of ELBO) was computed using the values before the rescaling step of the adaptive normalisation layer. This approach is also equivalent to applying an instance-level normalisation after the final output to calculate the error. Otherwise, distribution shifts at the patient level would lead to significant variations in the loss magnitude, making it impossible to find a unified threshold for identifying artefacts. Second, the accumulation method of loss matters. Specifically, the sum-based loss was employed instead of the mean-based approach to ensure sufficient gradient magnitudes for effective backpropagation. We empirically found that the traditional learning rate annealing method in VAE is insufficient to address this issue. This issue can lead to the ELBO loss approaching zero due to an imbalanced KL-divergence term, also known as posterior collapse, a common issue in variational autoencoder. Collapse often leads to the decoder generating only noise instead of meaningful physiological patterns. To prevent this, the decoder's reconstruction performance on the validation set was regularly inspected during training, ensuring the network learned physiologically relevant features.

After addressing these issues, the model was trained using the Adam optimiser with a learning rate of 0.001 and a batch size of 32. The training process was conducted over 300 epochs, employing early stopping with a patience parameter set to 50 epochs. This approach helps prevent overfitting by halting the training if the validation performance does not improve for 50 consecutive epochs. Data processing and model training were performed using an Intel Xeon Gold 6148 CPU (2.40GHz) and an NVIDIA GeForce RTX 3090, with Python 3.9.18.

**Baseline Models**

OCSVM[23]: A widely used unsupervised learning method in this field that identifies anomalies by maximising the boundary around normal examples and classifying deviations as artefacts. We use the same internal training set in our case to reproduce this algorithm.

XGBoost[25]: Highly popular supervised machine learning model, renowned for its gradient-boosting framework and exceptional performance across diverse datasets. To reproduce this algorithm, we use a 5-fold cross-validation with an 80-20 train test split in our held-out test set.

ResNet 1D[24]: A deep learning model tailored for one-dimensional time series data, which leverages residual connections to capture intricate temporal dependencies and has become a standard in physiological signal processing tasks. To reproduce this algorithm, we use a 5-fold cross-validation with an 80-20 train test split in our held-out test set.

**Arterial Blood Hypertension Events**

Hypertension is one of the most common and fatal symptoms associated with ABP and poses a significant global health risk. In this study, we adopt the WHO criteria for defining a hypertension event in our calculation: either a diastolic blood pressure (DBP, corresponding to the waveform's trough) > 90 mmHg or a systolic blood pressure (SBP, corresponding to the waveform's peak) > 140 mmHg. Although different diseases may have varying thresholds, this definition effectively demonstrates the impact of artefact cleaning. To identify these events, we first use a peak detection algorithm (via Scipy) to locate positive peaks corresponding to SBP, and then apply the same method to the inverted signal to detect negative peaks for DBP. During the artefact identification phase, segments contaminated by artefacts are marked as "Not a Number" (NaN) instead of 0, as 0 has physiological significance. After removing artefacts, the same detection method is applied to identify hypertension events in the cleaned data. This process allows us to compare the frequency and characteristics of hypertension events across segments with and without artefact contamination.

**Real-time Implementation**

The real-time implementation of our algorithm was integrated into the ICM+ Microsoft Windows-based software platform, running on an Intel Xeon CPU E5-2620 v3@ 2.40 GHz. The algorithm was first packaged as a plugin within the software, enabling activation for real-time performance evaluation. Using a simulated data stream, we measured the performance of the computer both with and without the plugin. For measuring the processing time, the plugin recorded the computation time required for a 10-second data sample. Simultaneously, we monitored resource utilisation, including CPU and memory usage, to evaluate the algorithm's computational overhead. This ensured that the algorithm met real-time requirements without compromising the responsiveness of the host computer.

# Code availability

The data preprocessing scripts, including the HDF5 data reader and our network code, are available at https://github.com/xuhang2019/GenClean.